# Analytical Modeling of Parameter Imbalance in Permanent Magnet Synchronous Machines

**Prerit Pramod**, *Senior Member*, IEEE

Control Systems Engineering, MicroVision, Inc.

*Email*: preritpramod89@gmail.com; preritp@umich.edu; prerit_pramod@microvsion.com

**Abstract –** This paper presents a systematic and comprehensive analysis of the impact of parameter imbalance in permanent magnet synchronous machines. Analytical models that reveal the effects of imbalance are obtained for each parameter. Thereafter, the models are verified for accuracy by comparison with complex simulations that closely represent true machine behavior. Such models may be utilized for developing (general) algorithms for detection, learning and mitigation of the negative effects of parameter imbalance including current (and thus torque) pulsations during real-time operation.

## Background

Industrial applications such as electric power steering (EPS) [1]–[5] that involve mass manufacturing of electric machines, including permanent magnet synchronous machines (PMSM) [6], [7], switched reluctance machines (SRM) [8]–[12], and permanent magnet DC machines (PMDC) [13] must maintain tight control over the part-to-part variation as well as intra-part balance of machine parameters. However, very tight control of such variations and imbalances is not practical since it results in high volume rejection of manufactured parts and thus unnecessary costs. Imbalance of machine parameters results in non-ideal current and thus torque control, i.e., undesirable current and torque pulsations are observed. This effect is significantly magnified when feedforward current control [14]–[19] is employed as opposed to feedback control [20]–[29], although even the latter suffers from this situation due to bandwidth and maximum bus voltage limitations. While the effect of parameter imbalance is somewhat understood, a detailed analysis of the same is still lacking.

A systematic and comprehensive analysis of the impact of parameter imbalance in PMSMs is presented here. Analytical (mathematical) models that reveal the effects of imbalance are obtained for each parameter. Such mathematical models expand the ability to capture mathematically non-ideal behavior that are typically not included in conventional formulations [30], [31]. Thereafter, the models are verified for accuracy by comparison with simulations that closely represent true machine behavior. Such models may be utilized for developing (general) algorithms for detection, learning and mitigation of the negative effects of parameter imbalance including current (and thus torque) pulsations during real-time operation [32]–[35]. Note that the focus of this write-up is on modeling of the actual machine. The behavior of the motor drive system during actual operation, where the motor control system interacts with the electric machine, is not presented here.

## Description

The mathematical model of a 3-phase PMSM in the stationary or abc reference frame consists of the electrical and magnetic relationships, i.e., the voltage to current relationship and the current to torque expression respectively. The electrical circuit equations are expressed as follows.

$$
\begin{aligned}
V_a &= R_a I_a + \dot{\lambda}_a \\
V_b &= R_b I_b + \dot{\lambda}_b \\
V_c &= R_c I_c + \dot{\lambda}_c \\
\lambda_a &= L_a I_a - M_{ab} I_b - M_{ac} I_c - \lambda_{am} \cos\theta \\
\lambda_b &= L_b I_b - M_{ba} I_a - M_{bc} I_c - \lambda_{bm} \cos(\theta - \beta) \\
\lambda_c &= L_c I_c - M_{ca} I_a - M_{cb} I_b - \lambda_{cm} \cos(\theta - 2\beta)
\end{aligned} \tag{1}
$$

where $V_x$ and $I_x$ are the phase voltages and currents for phase $x$, $R_x$, $L_x$ and $\lambda_{xm}$ are the phase resistance, self-inductance and permanent magnet flux linkage respectively, and $M_{xy}$ represents the mutual inductance of phase $x$ due to current in phase $y$. $\beta$ is the spatial angle difference between the different phases of the electric machine and is equal to $\frac{2\pi}{n}$ with $n$ being the number of phases. The electromagnetic torque is obtained from the current and flux linkages as follows.

$$
\begin{aligned}
T_e &= \frac{\partial W'}{\partial \theta} \\
W' &= \sum_{x=a,b,c} \int \lambda_x \, dI_x
\end{aligned} \tag{2}
$$

where $T_e$ represents the electromagnetic torque, $W'$ is the magnetic co-energy while $\theta$ is the electrical (phase) position of the motor. Thus, for modeling the mismatch or imbalance between phases, parameters may be written as follows.

$$
\begin{aligned}
R_x &= R + \Delta R_x \\
L_x &= L + \Delta L_x \\
M_{xy} &= M + \Delta M_{xy} \\
\lambda_{xm} &= \lambda_m + \Delta \lambda_x
\end{aligned} \tag{3}
$$

where the $\Delta A_x$ term represents the deviation of the value of parameter $A$ for phase $x$ from the nominal value $A_r$. For mathematical convenience, the lowest out of the parameter values between all the phases may be chosen to be the nominal value. In this way, the one of the error terms is always zero. The individual error terms may then be obtained by averaging the deviation of the individual phase parameters from the nominal value.

In general, the phase voltage equations are converted to the synchronously rotating or dq reference frame using the commonly known Clarke and Park transforms, which are expressed (in combined form) as follows.

$$
h_{dq0} = T_f h_{abc} \tag{4}
$$

$$T_f = \frac{2}{3}\begin{bmatrix} \cos\theta & \cos(\theta-\beta) & \cos(\theta-2\beta) \\ \sin\theta & \sin(\theta-\beta) & \sin(\theta-2\beta) \\ \frac{1}{2} & \frac{1}{2} & \frac{1}{2} \end{bmatrix}$$

where $h$ may represent the voltage or current. The inverse Clarke and Park transforms (again in combined form) are expressed as follows.

$$h_{abc} = T_i h_{dq0}$$
$$T_i = T^{-1} = \begin{bmatrix} \cos\theta & \sin\theta & 1 \\ \cos(\theta-\beta) & \sin(\theta-\beta) & 1 \\ \cos(\theta-2\beta) & \sin(\theta-2\beta) & 1 \end{bmatrix} \quad (5)$$

With matched or equal phase parameters, the Park transform results in machine equations that are independent of position. These ideal equations are commonly used for the purposes of modeling, estimation and control in most industrial motor drive control systems. In order to obtain the analytical model, all the parameters are assumed to be different (as explained above). The general phase voltage equations are then transformed into the dq frame utilizing the transformation matrices. This results in the following voltage equations.

$$\begin{aligned} V_d &= V_{di} + \Delta V_{dR} + \Delta V_{d\lambda} + \Delta V_{dLM} \\ V_q &= V_{qi} + \Delta V_{qR} + \Delta V_{q\lambda} + \Delta V_{qLM} \end{aligned} \quad (6)$$

where the subscript $i$ represents ideal (position independent) equations. The additional voltage terms, referenced by $\Delta V$, are obtained by applying the transformation considering the error terms due to the imbalance. The individual voltage terms that arise due to resistance, permanent magnet flux linkage and inductance imbalance are represented by subscripts $R$, $\lambda$ and $LM$ respectively. The derivation for obtaining these terms for each parameter individually is presented in the following description. The ideal dq frame model for non-salient pole machines is specified below.

$$\begin{aligned} V_{di} &= RI_d + (L+M)\left(\dot{I}_d + \omega_e I_q\right) \\ V_{qi} &= RI_q + (L+M)\left(\dot{I}_q - \omega_e I_d\right) + \omega_e \lambda_m \\ T_e &= \frac{3}{2}\frac{N_p}{2}\lambda_m I_q \end{aligned} \quad (7)$$

The ideal dq model considering salient pole machines consists of separate d and q axis inductances and is specified here for reference as follows.

$$\begin{aligned} V_{di} &= RI_d + \omega_e L_q I_q + L_d \dot{I}_d \\ V_{qi} &= RI_q - \omega_e L_d I_d + L_q \dot{I}_q + \omega_e \lambda_m \\ T_e &= \frac{3}{2}\frac{N_p}{2}\left(\lambda_m + \left(L_q - L_d\right)I_d\right)I_q \end{aligned} \quad (8)$$

Note that the torque expressions for modeling imbalance of different parameters are not shown here. However, they can be easily obtained by following the same idea as the voltage-current derivations. It is important to understand that the models presented here are general (plant) models that describe the machine behavior and are not influenced by the control strategy whatsoever. Further, the models are valid for all synchronous machines, including wound-rotor machines with field current windings.

## Resistance Imbalance

The additional voltage terms obtained as a result of resistance imbalance are specified below.

$$\frac{3}{2}\Delta V_{dR} = \Delta R_a I_a \cos\theta + \Delta R_b I_b \cos(\theta - \beta) + \Delta R_c I_c \cos(\theta - 2\beta)$$

$$\frac{3}{2}\Delta V_{qR} = \Delta R_a I_a \sin\theta + \Delta R_b I_b \sin(\theta - \beta) + \Delta R_c I_c \sin(\theta - 2\beta)$$

$$\begin{gathered} \Delta V_{dR} = \frac{\Delta R}{3} I_d + K_R \cos(2\theta + \phi_R) I_d + K_R \sin(2\theta + \phi_R) I_q + (\dots) I_0 \\ \Delta V_{qR} = K_R \sin(2\theta + \phi_R) I_d - K_R \cos(2\theta + \phi_R) I_q \\ K_R = \frac{1}{3}\sqrt{\Delta R_a^2 + \Delta R_b^2 + \Delta R_c^2 - \Delta R_a \Delta R_b - \Delta R_b \Delta R_c - \Delta R_c \Delta R_a} \\ \phi_R = \tan^{-1}\left(\frac{\sqrt{3}(-\Delta R_b + \Delta R_c)}{2\Delta R_a - \Delta R_b - \Delta R_c}\right) \end{gathered} \tag{9}$$

A block diagram representation of the effect of resistance imbalance is shown in the figure below.

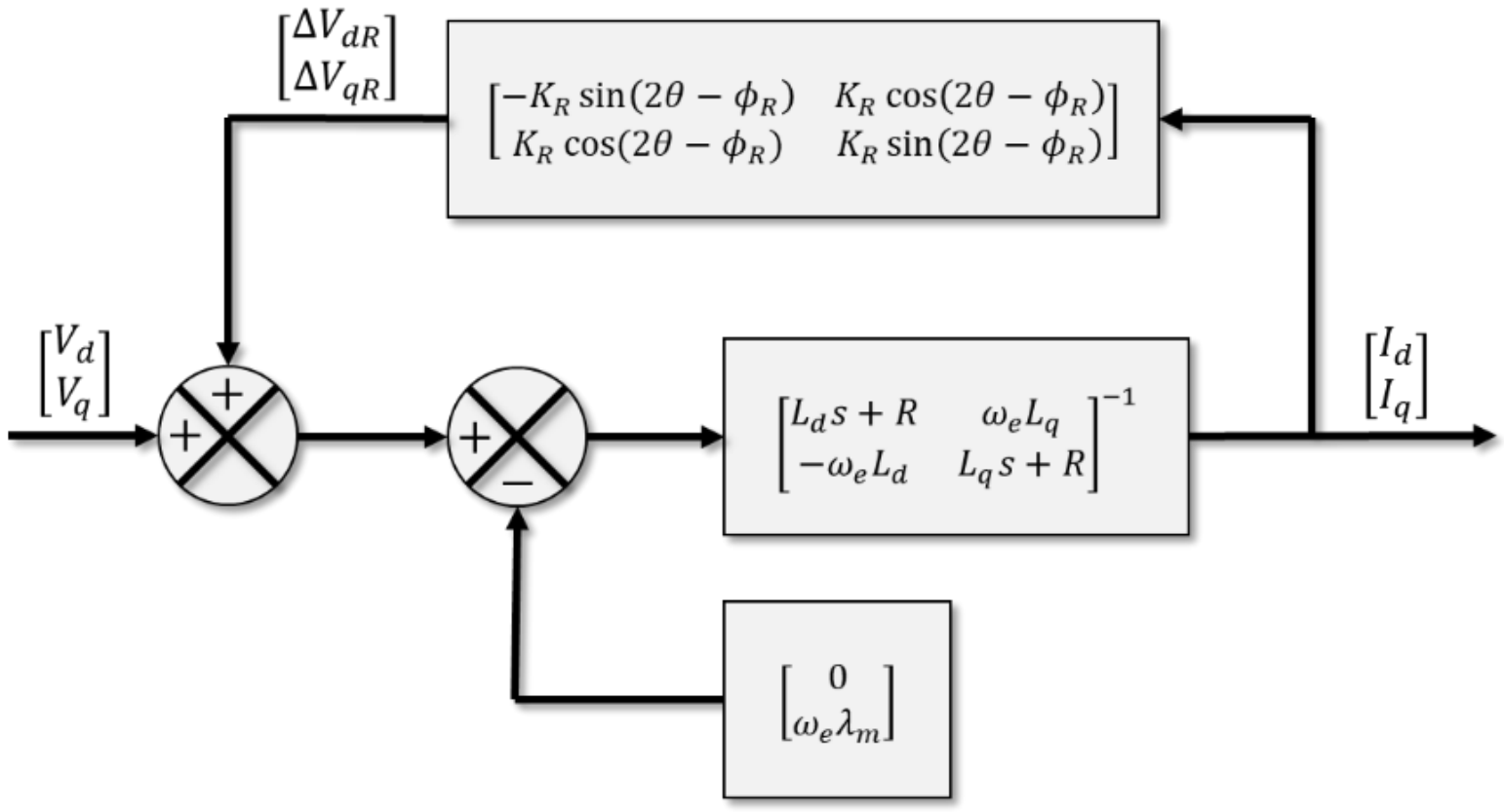

*Figure 1*: Block diagram representation of analytical model for resistance imbalance.

A comparison of the analytical prediction of resistance imbalance with a detailed simulation model having high accuracy for describing true machine behavior is shown in the figure below.

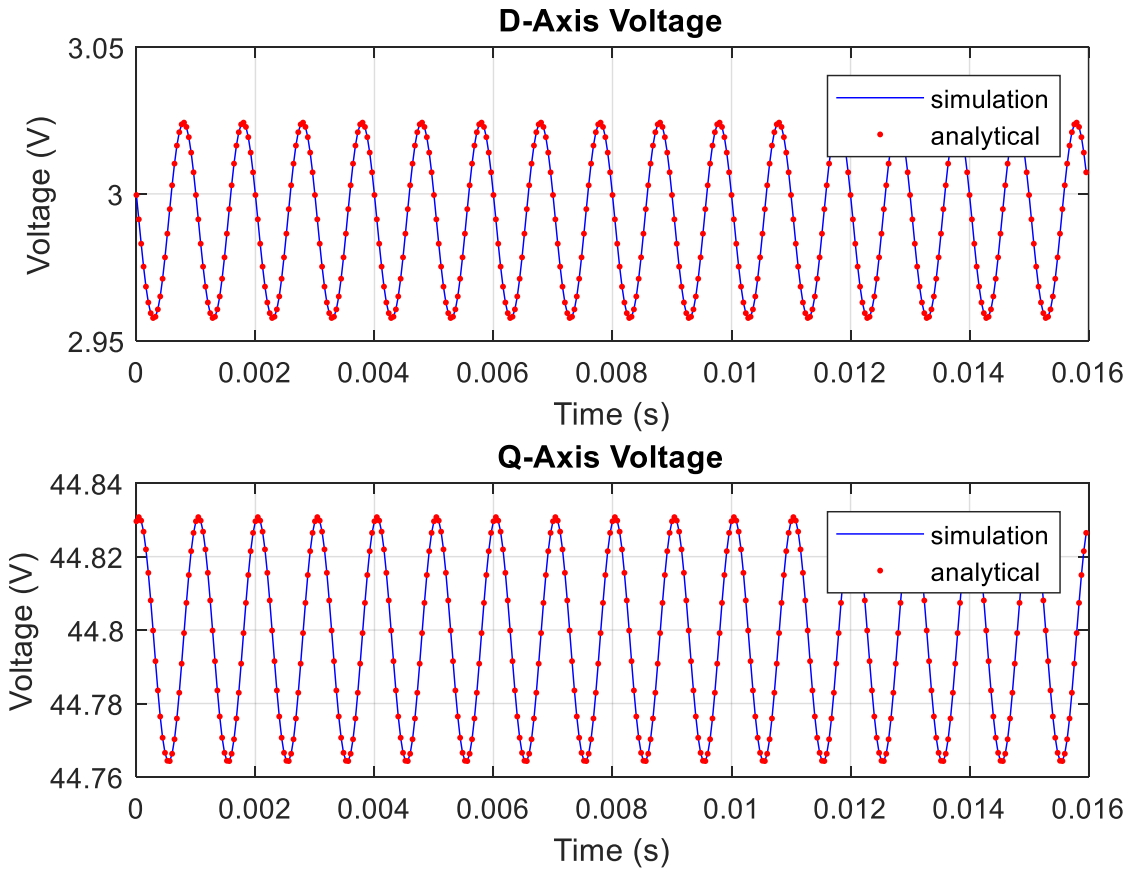

*Figure 2*: Results illustrating accuracy of analytical model for resistance imbalance.

### Permanent Magnet Flux Linkage Imbalance

The additional voltage terms obtained as a result of permanent magnet flux linkage imbalance are as follows.

$$\frac{3}{2}\Delta V_{d\lambda} = \omega_e \Delta\lambda_{am} \sin\theta\cos\theta + \omega_e \Delta\lambda_{bm} \sin(\theta-\beta)\cos(\theta-\beta) + \omega_e \Delta\lambda_{cm} \sin(\theta-2\beta)\cos(\theta-2\beta)$$

$$\frac{3}{2}\Delta V_{q\lambda} = \omega_e \Delta\lambda_{am} \sin^2\theta + \omega_e \Delta\lambda_{bm} \sin^2(\theta-\beta) + \omega_e \Delta\lambda_{cm} \sin^2(\theta-2\beta)$$

$$\Delta V_{d\lambda} = \omega_e K_\lambda \sin(2\theta+\phi_\lambda)$$
$$\Delta V_{q\lambda} = \omega_e(\Delta\lambda_{am}+\Delta\lambda_{bm}+\Delta\lambda_{cm}) - \omega_e K_\lambda \cos(2\theta+\phi_\lambda) \quad (10)$$
$$K_\lambda = \frac{1}{3}\sqrt{\Delta\lambda_a^2+\Delta\lambda_b^2+\Delta\lambda_c^2-\Delta\lambda_a\Delta\lambda_b-\Delta\lambda_b\Delta\lambda_c-\Delta\lambda_c\Delta\lambda_a}$$
$$\phi_\lambda = \tan^{-1}\left(\frac{\sqrt{3}(-\Delta\lambda_b+\Delta\lambda_c)}{2\Delta\lambda_a-\Delta\lambda_b-\Delta\lambda_c}\right)$$

A block diagram representation of the effect of permanent magnet flux linkage imbalance is as follows.

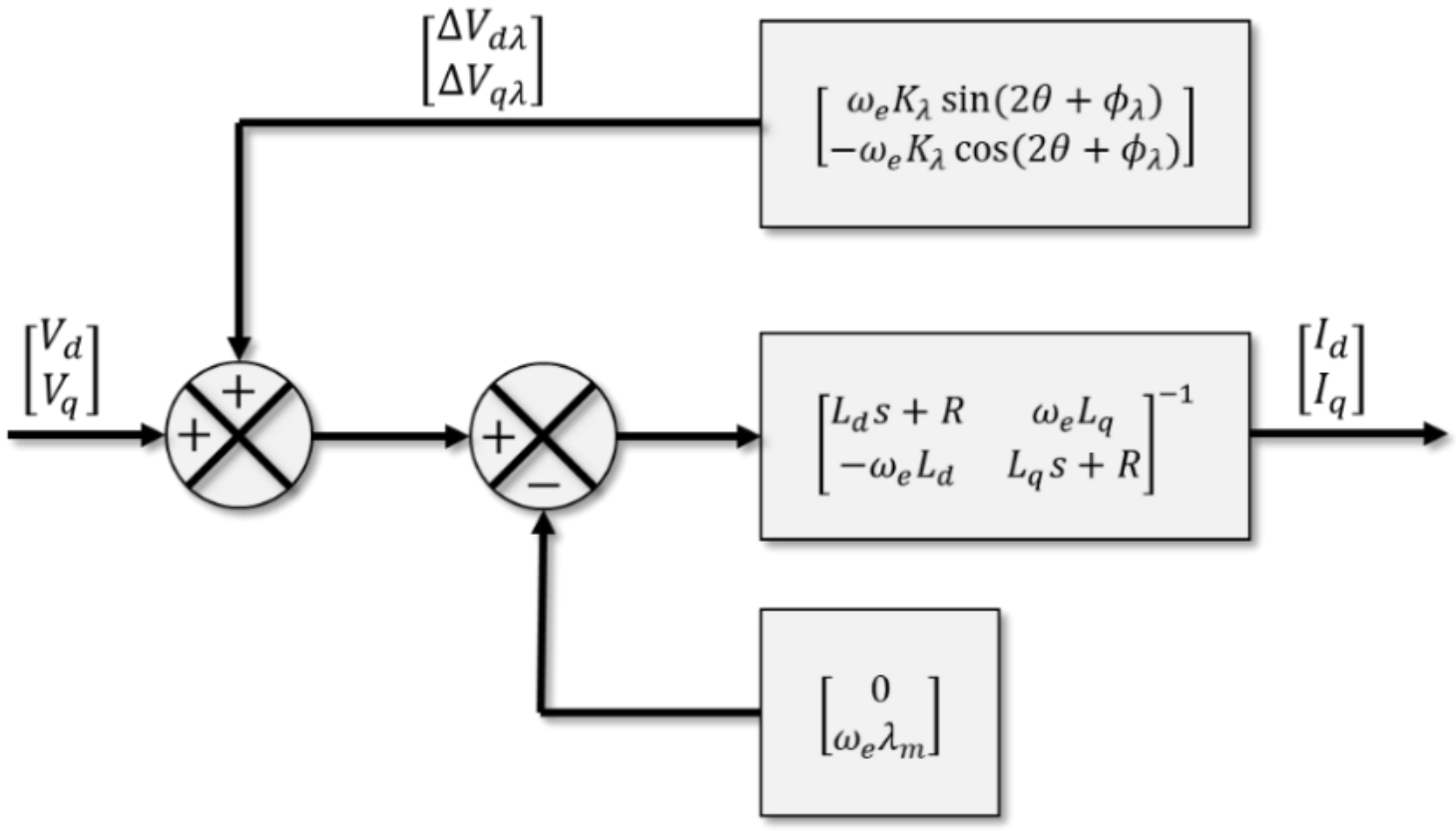

*Figure 3*: Block diagram representation of analytical model for permanent magnet flux linkage imbalance.

A comparison of the analytical prediction of permanent magnet flux linkage imbalance with a detailed simulation model having high accuracy for describing true machine behavior is shown in the figure below.

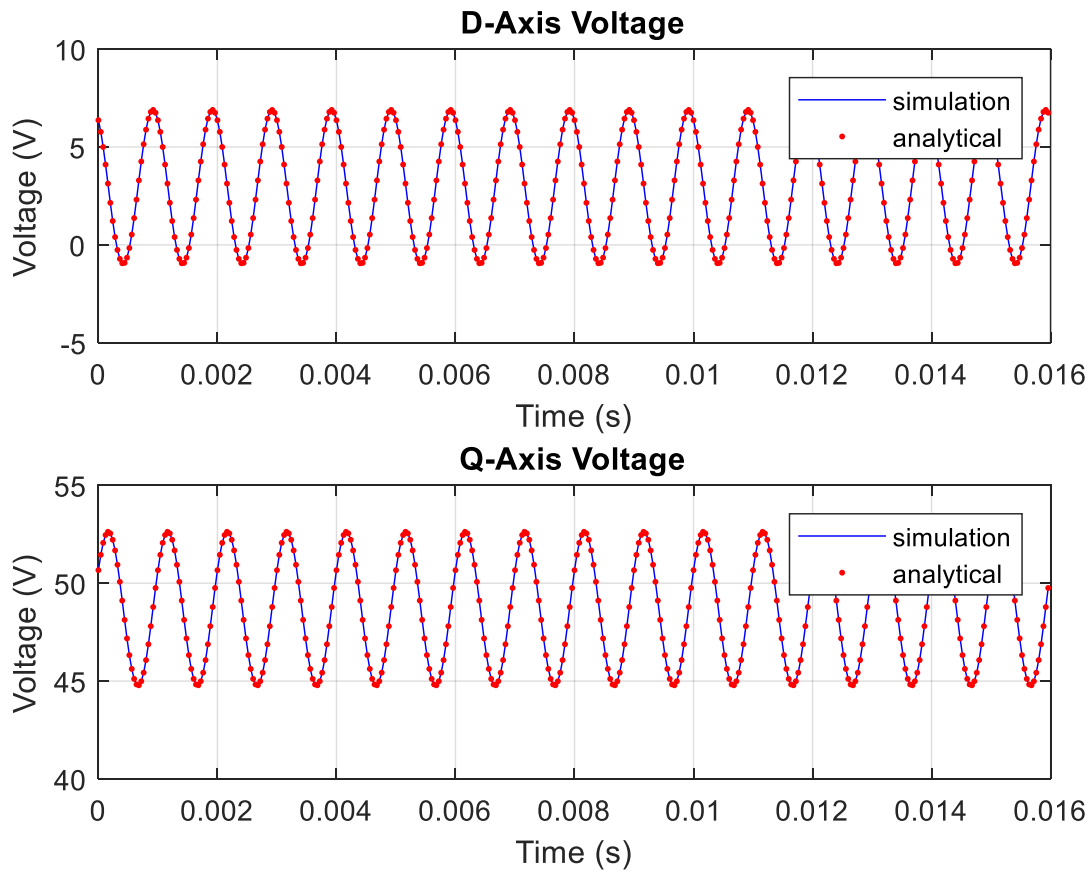

*Figure 4*: Results illustrating accuracy of analytical model for permanent magnet flux linkage imbalance.

## Inductance Imbalance

The additional voltage terms obtained as a result of inductance (including both self and mutual inductances) imbalance are specified below.

$$
\begin{aligned}
\frac{3}{2}\Delta V_{dL} &= \big(p(L_a I_a - M_{ab} I_b - M_{ac} I_c)\big)\cos\theta + \big(p(L_b I_b - M_{ba} I_a - M_{bc} I_c)\big)\cos(\theta-\beta) + \big(p(L_c I_c - M_{ca} I_a - M_{cb} I_b)\big)\cos(\theta-2\beta) \\
\frac{3}{2}\Delta V_{qL} &= \big(p(L_a I_a - M_{ab} I_b - M_{ac} I_c)\big)\sin\theta + \big(p(L_b I_b - M_{ba} I_a - M_{bc} I_c)\big)\sin(\theta-\beta) + \big(p(L_c I_c - M_{ca} I_a - M_{cb} I_b)\big)\sin(\theta-2\beta)
\end{aligned} \tag{11}
$$

where $p$ represents the derivative operator. This is the general expression for all permanent magnet synchronous machines (PMSMs). In the case of salient pole PMSMs both the self and mutual inductance terms are position dependent and so the derivative operation needs to be carried out accordingly. For non-salient pole machines, the inductances may be assumed to be position independent.

$$
\begin{aligned}
\Delta V_{dL} &= (\Delta L + \Delta M + K_L\cos(2\theta+\phi_L) - K_M\cos(2\theta+\phi_M))(\dot{I}_d + \omega_e I_q) + (K_L\sin(2\theta+\phi_L) - K_M\sin(2\theta+\phi_M))(-\omega_e I_d + \dot{I}_q) \\
\Delta V_{qL} &= (K_L\sin(2\theta+\phi_L) - K_M\sin(2\theta+\phi_M))(\dot{I}_d + \omega_e I_q) + (\Delta L + \Delta M + K_L\cos(2\theta+\phi_L) + K_M\cos(2\theta+\phi_M))(-\omega_e I_d + \dot{I}_q)
\end{aligned}
$$

$$
\begin{aligned}
K_L &= \frac{1}{3}\sqrt{\Delta L_a^2 + \Delta L_b^2 + \Delta L_c^2 - \Delta L_a\Delta L_b - \Delta L_a\Delta L_c - \Delta L_b\Delta L_c} \\
\phi_L &= \tan^{-1}\left(\frac{\sqrt{3}(-\Delta L_b + \Delta L_c)}{2\Delta L_a - \Delta L_b - \Delta L_c}\right) \\
K_M &= \frac{2}{3}\sqrt{M_{ab}^2 + M_{bc}^2 + M_{ca}^2 - M_{ab}M_{ac} - M_{ab}M_{cb} - M_{ac}M_{cb}} \\
\phi_M &= \tan^{-1}\left(\frac{\sqrt{3}(-M_{ab} + M_{ac})}{-M_{ab} - M_{ac} + 2M_{cb}}\right)
\end{aligned} \tag{12}
$$

A block diagram representation of the effect of inductance imbalance in shown in the figure below.

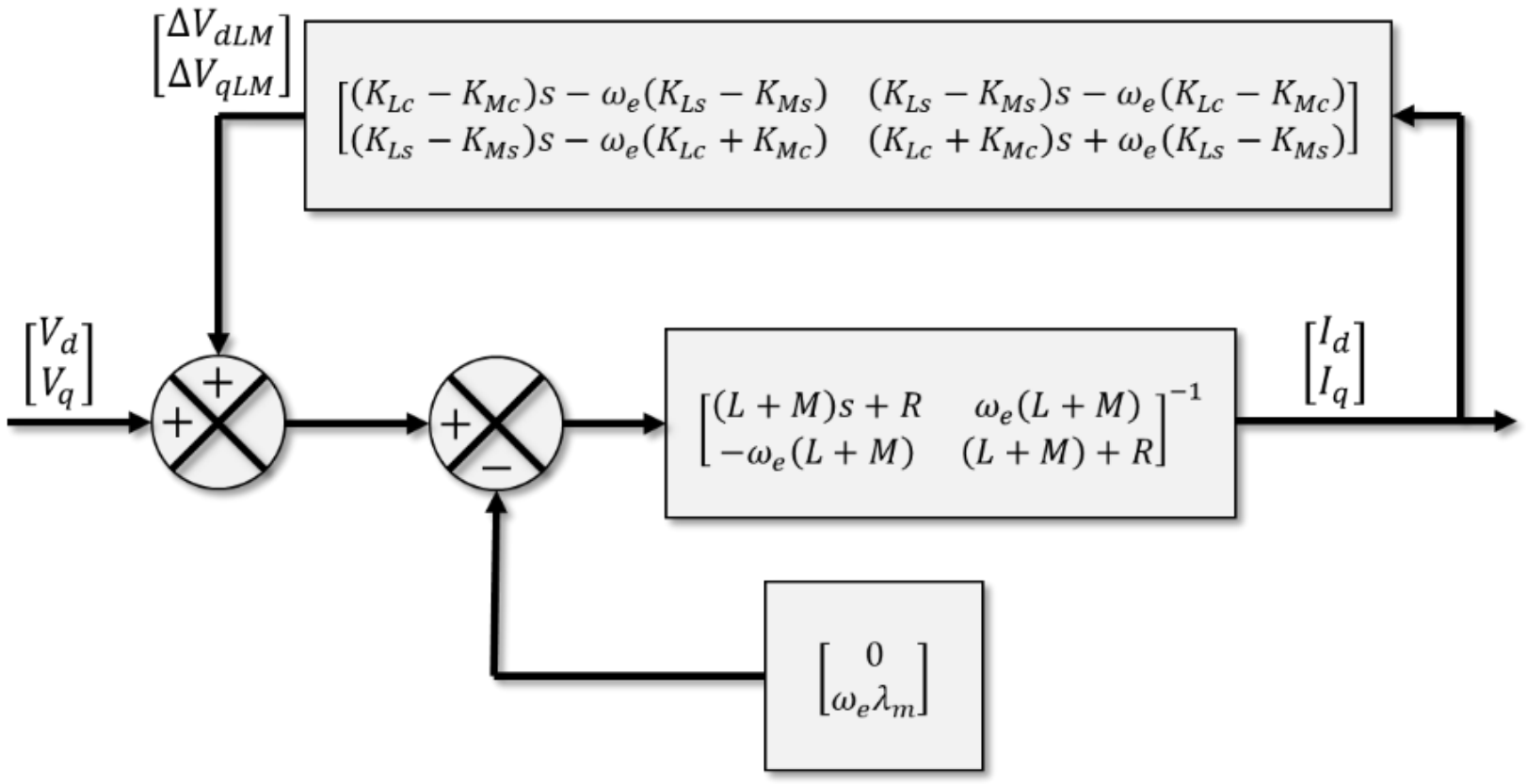

*Figure 5*: Block diagram representation of analytical model for inductance imbalance.

A comparison of the analytical prediction of inductance imbalance with a detailed simulation model having high accuracy for describing true machine behavior is shown in the figure below.

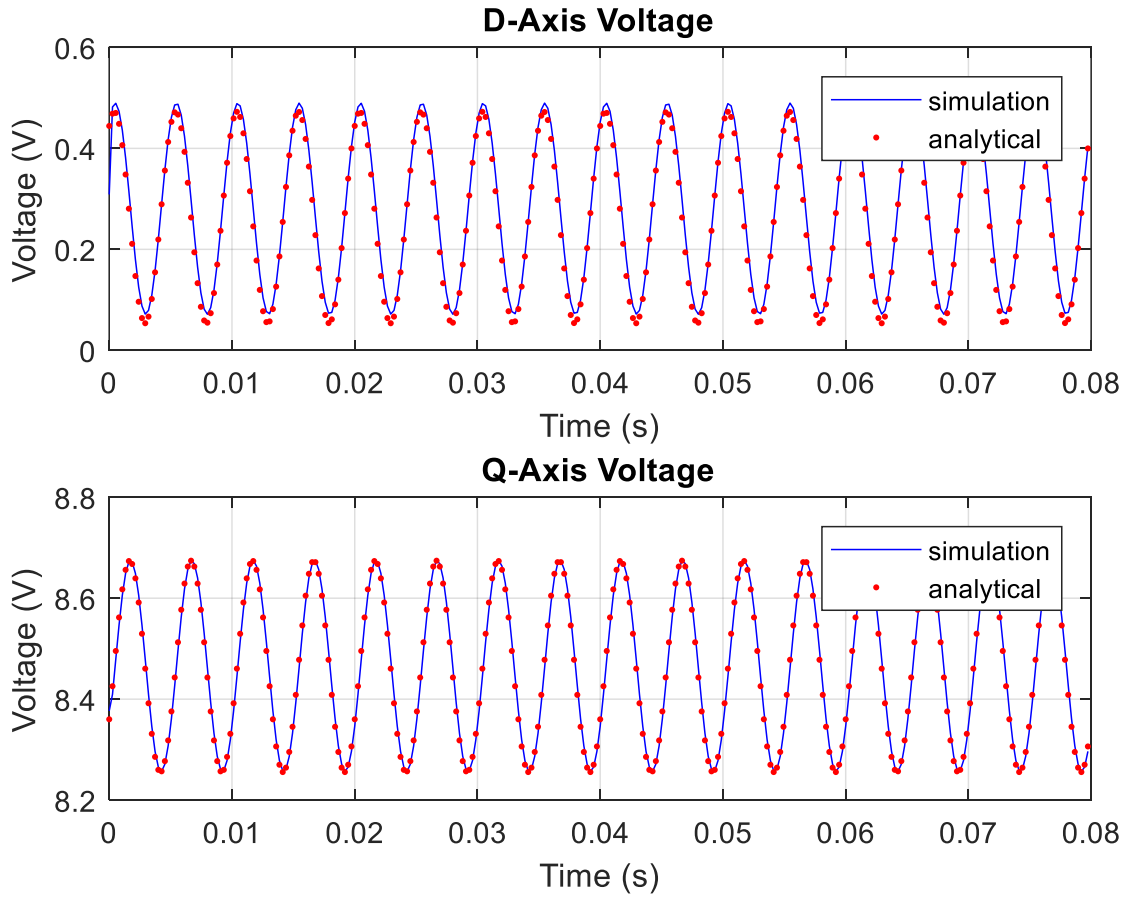

*Figure 6*: Results illustrating accuracy of analytical model for inductance imbalance.

Note that the above derivations concerning inductance imbalance are only valid for non-salient pole PMSMs. While the derivation or results for modeling salient pole machines are not shown here, it is easy to extend the idea presented here to obtain those as well. As mentioned earlier, for salient pole machines, additional terms will be introduced due to the existence of second order position dependent terms in the stationary frame self and mutual inductances and therefore the derivative operator must be applied appropriately to correct determine the desired inductance imbalance model for salient pole synchronous machines.

## Conclusions

This paper presents analytical models capturing the effects of imbalance for all the different parameters of PMSMs are presented. These models are not commonly known and may be used to develop algorithms (that may be implemented at the manufacturing end of line or in the controller software for real-time operation) for the detection, identification, learning and mitigation of the negative effects of parameter imbalance in PMSM machines.